\documentclass[conference]{IEEEtran}
\IEEEoverridecommandlockouts
\usepackage{cite}
\usepackage{amsmath,amssymb,amsfonts}
\usepackage{algorithmic}
\usepackage{graphicx}
\usepackage{textcomp}
\usepackage{xcolor}
\usepackage{float,subfig}
\begin{document}

\title{Memristive Learning Cellular Automata: Theory and Applications\\
}

\author{\IEEEauthorblockN{Rafailia-Eleni Karamani\IEEEauthorrefmark{1},
Iosif-Angelos Fyrigos\IEEEauthorrefmark{1},
Vasileios Ntinas\IEEEauthorrefmark{1}\IEEEauthorrefmark{2},
Orestis Liolis\IEEEauthorrefmark{1},\\
Giorgos Dimitrakopoulos\IEEEauthorrefmark{1},
Mustafa Altun\IEEEauthorrefmark{3},
Andrew Adamatzky\IEEEauthorrefmark{4}, 
Mircea R. Stan\IEEEauthorrefmark{5},
Georgios Ch. Sirakoulis\IEEEauthorrefmark{1}, }
\IEEEauthorblockA{\IEEEauthorrefmark{1}Department of Electrical and Computer Engineering, Democritus University of Thrace, Xanthi, Greece}
\IEEEauthorblockA{\IEEEauthorrefmark{2}Department of Electronic Engineering, Universitat Polyt\'ecnica de Catalunya, Barcelona, Spain}
\IEEEauthorblockA{\IEEEauthorrefmark{3}Department of Electronics and Communication Engineering, Istanbul Technical University, Istanbul, Turkey}
\IEEEauthorblockA{\IEEEauthorrefmark{4}Unconventional Computing Centre, University of the West of England, Bristol, United Kingdom}
\IEEEauthorblockA{\IEEEauthorrefmark{5}Department of Electrical and Computer Engineering, University of Virginia, Charlottesville, Virginia, USA}}



\maketitle

\begin{abstract}
Memristors are novel non volatile devices that manage to combine storing and processing capabilities in the same physical place. Their nanoscale dimensions and low power consumption enable the further design of various nanoelectronic processing circuits and corresponding computing architectures, like neuromorhpic, in memory, unconventional, etc.
%
One of the possible ways to exploit the memristor’s advantages is by combining them with Cellular Automata (CA). CA 
constitute a well known non von Neumann computing architecture that is based on the local interconnection of simple identical cells forming $N$-dimensional grids. These local interconnections allow the emergence of global and complex phenomena. In this paper, we propose a hybridization of the CA original definition coupled with memristor based implementation, and, more specifically, we focus on Memristive Learning Cellular Automata (MLCA), which have the ability of learning using also simple identical interconnected cells and taking advantage of the memristor devices inherent variability. The proposed MLCA circuit level implementation is applied on optimal detection of edges in image processing through a series of SPICE simulations, proving its robustness and efficacy. 
\end{abstract}


\section{Introduction}
In 1971, Prof. L. Chua postulated the existence of a passive two-terminal circuit element, named memristor \cite{chua1971memristor} during his research on non-linear circuit analysis theory. Symmetry reasons among the equations formed by the four fundamental circuit variables, namely, current $i$, voltage $v$, charge $q$ and flux $f$ was what led Prof. Chua to his theoretical discovery to introduce a non-linear mathematical relationship between charge and flux. 
From that day on and after the seminal, memristor related, experimental work, by the HP researchers \cite{strukov2008missing} in 2008, this discovery has triggered a vast number of 
memristor applications in various scientific fields \cite{vourkasbook,chua2019handbook}, such as neuromorphic computing \cite{wang2017memristors}, logic design \cite{papandroulidakis2014boolean,kvatinsky2013memristor}, memory design \cite{ho2010dynamical,hamdioui2014memristor}, etc.  Memristors are considered nanoelectronic low power devices, which are characterised by their capability of non volatile information storage combined with the potential to perform computations on the same device. Nevertheless, and towards their fabrication, intrinsic variability of memristor devices usually affects severely their efficacy to perform adequately in an always standardized manner. 

On the other hand, one of the most well-known computational architectures is 
Cellular Automata (CA), 
whose basic inherent characteristics include massive parallelism, local interactions, and not complicated topologies \cite{vonneumann1966theory}. Moreover, in the original definition of CA, memory and local processing rule are encapsulated in the same site, i.e. the CA cell, making CA rather suitable for in-memory computing \cite{sirakoulis2015robots}.
As such, when CAs are coupled with the memristor nanodevices, the resulting computation paradigm is expected highly auspicious. Furthermore, enrichment of the CA characteristics would be possible by allowing them to present learning capabilities to fully adapt to any random environment. This way, the CA can receive feedback from its past actions and learn how to improve its final output, aiming to an optimal goal and improvement of its performance. One way to achieve this is by merging CA with another class of automata structures, namely the Learning Automata (LA) \cite{narendra1974learning}. 

In this paper, we propose a Memristive Learning Cellular Automata (MLCA) architecture that combines the characteristics of CA, LA and memristors, in both theoretical and device/implementation level, aiming to advance the functionality of the CA paradigm with inherent learning features owing to memristor devices intrinsic variability. The proposed MLCA architecture is also stressed for various local rules aiming to tackle well defined existing problems in real life applications like image processing. In particular, MLCA circuits are applied to edge detection problem in an efficient way in terms of design complexity mainly attributed to CA grid, area utilization because of the same CA cell usage in the proposed design, low power consumption mainly due to memristor device properties, and overall self learning performance arriving from the memristor devices characteristics. 

\section{Brief Introduction on Automata Theoretical Principles}
\label{Section:Intro}
\subsection{Cellular Automata}
\label{Subsection:CA}
A CA is a computing architecture originally proposed by J. von Neumann and S. Ulam in the 1940's \cite{vonneumann1966theory} and more widely studied in the past decades by S. Wolfram \cite{wolfram2002new}. Their popularity is due to the fact that they allow the emergence of complex phenomena employing simple structures and resulting to inherent emergent computation and self-organization. 

A CA can be defined using specific attributes, as follows: 
\begin{itemize}
  \item It consists of a finite size $N$-dimensional grid of CA cells.
  \item Each of these CA cells can be in one of a predefined set $S$ of states (in this work, $S = {0,1}$).
  \item For every CA cell, the CA's neighbourhood is described as a set of attached cells connected to the CA cell, affecting its time evolution. 
  \item Every CA cell's next state is computed through a fixed cell state transition rule $F$, taking into account the cell's and its neighbours' current states.
\end{itemize}

 CA local interaction rule is considered identical for all CA cells, unless otherwise clarified in possible hybridisation of the CA definition, and it is applied in a fully synchronous manner in discrete time steps to every CA cell on the CA grid. In the case of two-dimensional ($2-D$) CA, the most well known neighbourhoods are (i) the von Neumann neighbourhood, where the state transition rule $F$ can be defined as follows, practically introducing the involvement of the cell states of the most nearest adjacent neighbours:
 
 \begin{equation}
 S^{\tau+1}_C=F\big(S^{\tau}_C,S^{\tau}_N,S^{\tau}_E,S^{\tau}_S,S^{\tau}_W\big),
 \label{eq1}
 \end{equation}
 
 \noindent while in case of (ii) 
 Moore neighbourhood, beyond the aforementioned adjacent cells, the state transition rule successfully involves also the states of the diagonal neighbouring cells:
 \begin{equation}
 S^{\tau+1}_C=F\big(S^{\tau}_C,S^{\tau}_N,S^{\tau}_E,S^{\tau}_S,S^{\tau}_W,S^{\tau}_{NE},S^{\tau}_{NW},S^{\tau}_{SE},S^{\tau}_{SW}\big).
 \label{eq2}
\end{equation}


\subsection{Learning Automata}

Learning Automata (LA) are a model of stochastic automata operating in random environments. In this class of automata, every action is selected based on an action probability vector which is updated, aiming to improve its performance.

The basic operation of a LA is as follows. Initially, there is no optimal action so all the actions have equal probabilities to occur. As the automaton interacts with its environment, an action is randomly selected and the probability for this action is rewarded or punished depending on the environment's response. So, the action probability vector is updated in every time step and another action is selected based on the updated action probability vector. This procedure is repeated until the automaton learns to choose the optimal action, which is the action with the highest probability to occur. All the above can be summed with the following equation: 

\begin{equation}
p(n+1) = T(p(n),a(n),x(n))
\label{eq3}
\end{equation}

\noindent where $p(n+1)$ is the action probability vector at the next time step, $p(n)$ is the current probability vector, $a(n)$ is the action selected by the automaton and $x(n)$ is the input to the automaton at a specific time step $n$. 


\subsection {Learning Cellular Automata}

Having explained in brief the basic notions of CA and LA, it is straightforward how these two models can be effectively coupled into Learning Cellular Automata (LCA). This architecture combines the parallel computation using local interactions and self organisation abilities of CA with the learning capabilities in an unknown environment of a LA. Therefore, the LCA is able to learn the optimal response of the environment taking into account feedback from its neighbours and itself. Implementing LCA requires having one LA in every CA cell. Interaction with its neighbourhood in every time step will modify the action probability vector accordingly so that each cell learns its optimal state.

\section{Memristor Basics}
The memristor is a nanoscale resistive device with the ability to remember its previous state. Its state can be affected either by the voltage applied to its terminals (voltage controlled) or the current passing through it (current controlled). In this work, the memristor model used is a voltage controlled, threshold based behavioural memristor model 
\cite{vourkas2015spice}. This model can be in either of 2 states, high resistance state ($R_{off}$) and low resistance state ($R_{on}$). 
The aforementioned characteristics will be utilized to meet with the requirements of the MLCA circuit design.

Moreover, one of the memristor's inherent characteristics is variability. Variability in a memristive device is defined as a large variance in the device's switching. The main reason for the existence of variability in memristive devices is that the fabrication 
techniques as well as the materials used are not yet mature enough to further ensure that all memristive devices will behave in the same way. The existence of variability is highly dependant on the thermodynamic properties of the selected materials \cite{kim2016voltage}. Some of the existing ways to help encounter this phenomenon include the appropriate selection of materials and external control of the circuit by applying voltage pulses of high enough amplitude to trigger the switching of the device to help tackle with the device's inability to switch \cite{yang2013memristive,kim2016voltage}. In this work, the inherent variability of the memristor is exploited to implement the stochastic behavior of the developed circuit.

\section{Design of a Memristive Learning Cellular Automaton Cell}
\begin{figure}[!t]
\centerline{\includegraphics[width=1\linewidth]{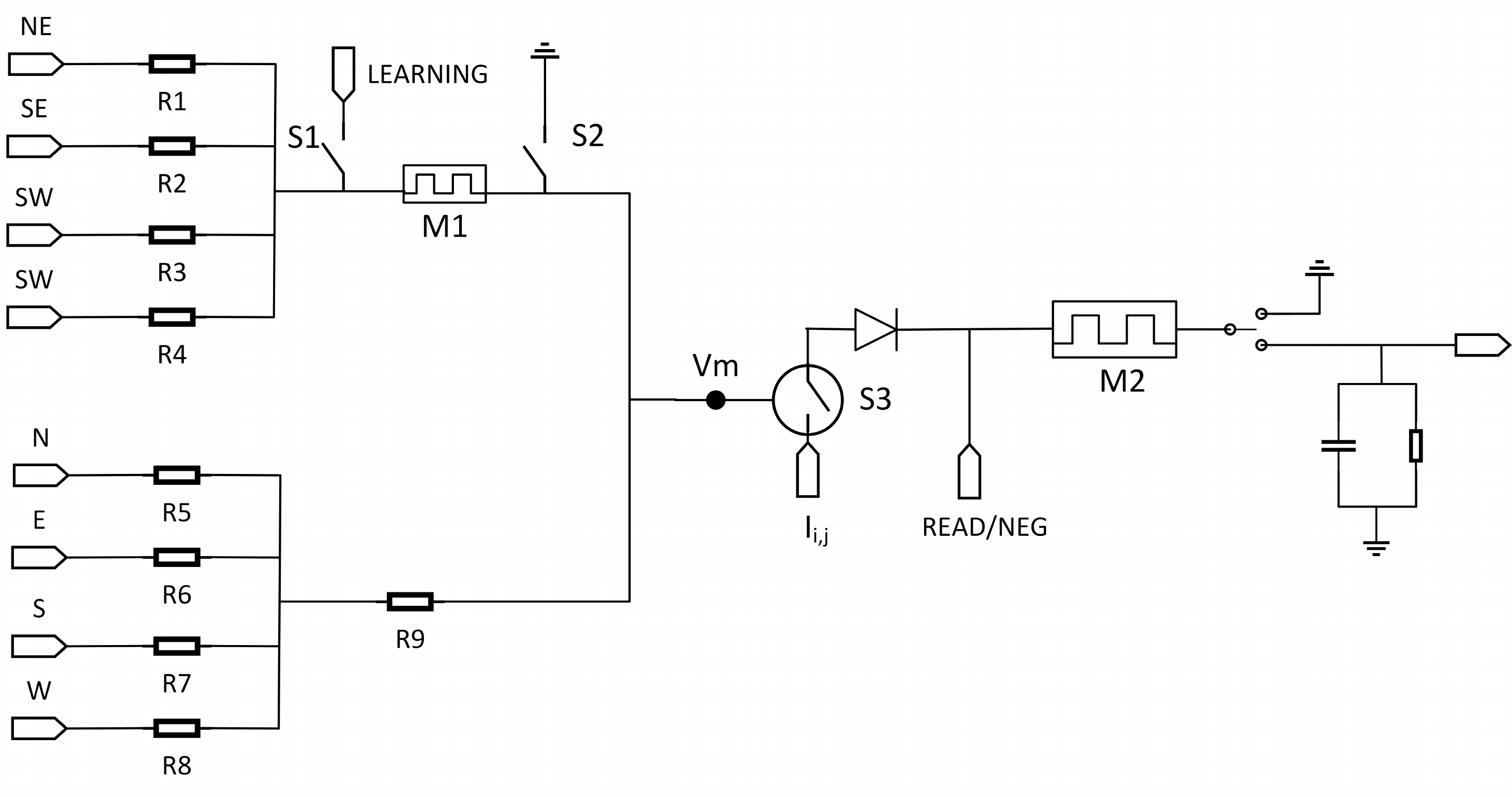}}
\caption{Schematic circuit representation of the MLCA cell for optimal detection of edges.}
\label{Cell}
\end{figure}

M. Itoh and L. Chua were the first referring to the theoretical principles of memristive CA \cite{Itoh2009} aiming to image processing applications. Other recent works have employed memristive CA for AI applications, $NP$-complete hard to be solved problems as well as epilepsy modeling \cite{vourkasbook,Stathisiet,adamatzky2011memristive,KaramaniISCAS,karamani2018game}. In this work, we  propose a circuit level implementation of a memristive cellular automaton that incorporates learning capabilities. 
%
%
Each time step is divided into two separate phases, the reading phase and the writing phase. The former is when a voltage pulse, whose amplitude is lower than the memristor's threshold, is applied to its first terminal while the other one is connected to the cell's output node in order to read its state and determine the cell's output. During the latter, each cell receives inputs from its neighbouring cells and according to the CA evolution rule, computes its next state. In this case, the memristor's second terminal is grounded so that the voltage applied to its first terminal can affect its state. Each time step's duration is selected to be constant
with the reading phase occupying a portion (1/10) of this duration and the rest being devoted for the computation of the cell's next state. The alternation between these two states is achieved using two voltage controlled switches 
(shown as a hybrid three (3) terminal switch in Fig.~\ref{Cell}) that is grounded during writing phase and connected to the output during reading phase.
The amplitude of the pulse generated during the reading phase has to be maintained high enough during the writing phase so as to be transmitted to the cell's neighbours and affect their states. For this reason, a resistor - capacitor pair with appropriate values is connected to every cell's output node 
thus allowing, through the slow discharge of the capacitor, to maintain the output voltage high to perform the next state's computation.
%
%

All the above described characteristics are crucial for storing the cell's state and transmitting it to its neighbours. In order to compute the cell's next state and incorporate the learning capabilities of LA, we will exploit the previously presented inherent variability in memristive devices. 
More specifically, we utilize this characteristic to embed the probability of action selection in every time step. Therefore, the probability of a device switching will be further enhanced if the cell's selection was right by applying a voltage high enough to ensure the device's switching. In the opposite case, the probability will be further reduced, and the amplitude of the voltage applied to the device will not guarantee its switching.

\section{Application of Memristive Cellular Learning Automata in Binary Image Edge Detection}

The architecture of MLCA aims to deliver promising results in figures of merits for various applications. 
In this paper, the proposed application example is image processing and specifically, optimal detection of edges in binary images, where each MLCA cell aims to learn which is the best neighbourhood selection for itself. Every cell in the 2-D CA grid represents an image pixel with two available neighbourhood arrangements, i.e. von Neumann and Moore. 
%
Initially, all cells will choose the neighbourhood type from the available set randomly, based on the probability vector. Every cell's next state will be computed based on the following rules. In binary images, where pixels can only be either 1 or 0, 
if the pixel is 1 and at least one pixel in the pixel's neighbourhood is 0, then it will be determined to be an edge, while if all the neighbouring pixels are 1, then the pixel is not identified as an edge. 
Otherwise, if the pixel's value is 0, it is never identified as an edge pixel.

The MLCA's next state now has 1's only to the cells identified as edges and we also need to calculate the reinforcement signal of the LA and update the probability vector. This process is repeated for a preselected number of time steps by always feeding the MLCA with the initial image and computing the new probability vector. The rule for updating the probability vector is that if all the MLCA in the same neighbourhood choose the same action, the probability for this action will be enhanced (reward) in the next calculation. Otherwise, it will be reduced (penalty). Therefore, the MLCA's goal is to gradually learn which neighbourhood type is appropriate, in particular segments of the image, and make a similar choice for all of them, converging to an optimal image map. 

\begin{figure}[!t]
\centering
\includegraphics[width=0.75\linewidth]{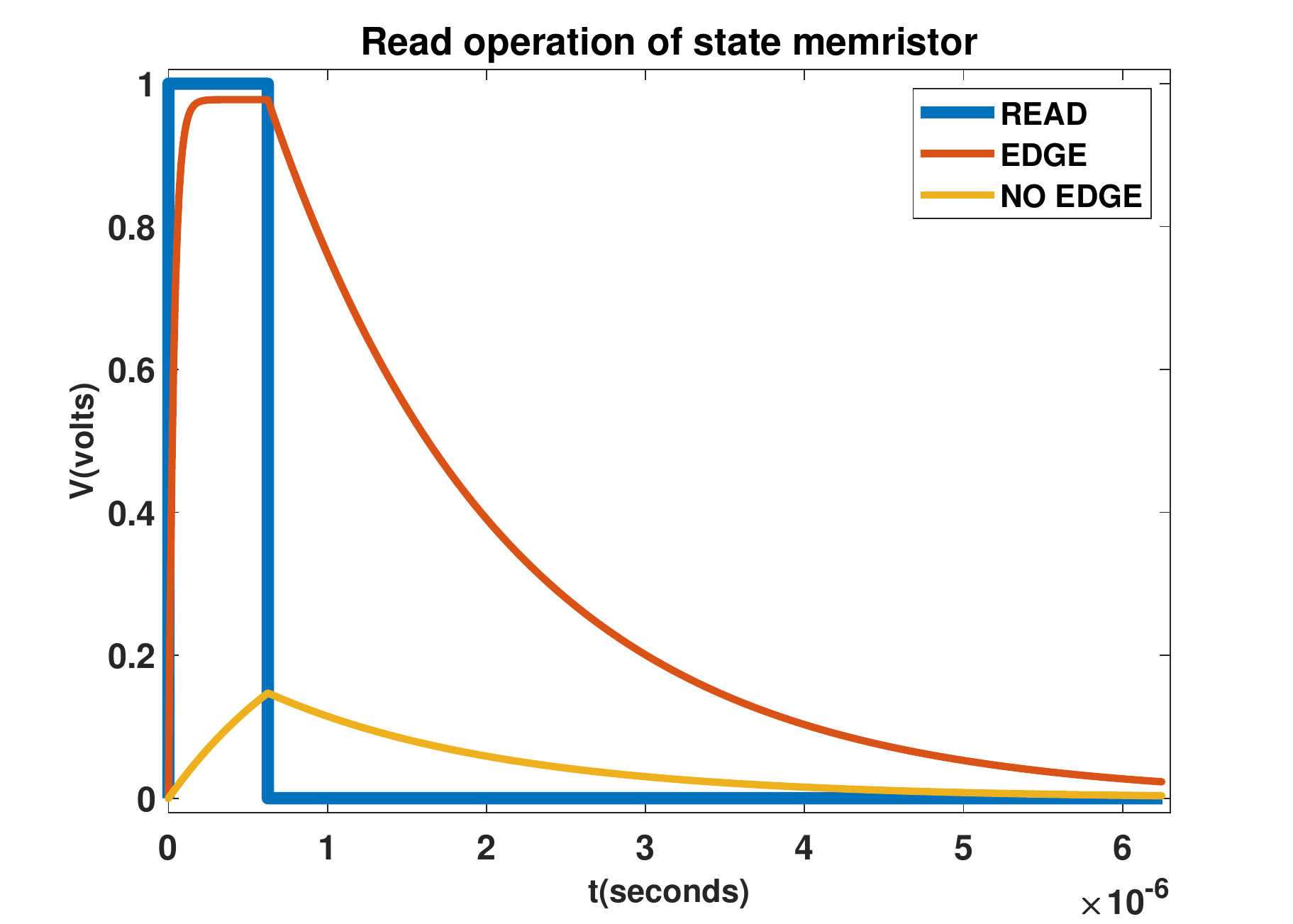}
\caption{Input READ Pulse and cell's output response for cell state $EDGE=1$ (red) and  $NO EDGE=0$ (yellow).}
\label{ON_OFF_response}
\end{figure}
  

\begin{figure*}[!t]
    \centering
  \subfloat{\includegraphics[width=0.45\textwidth]{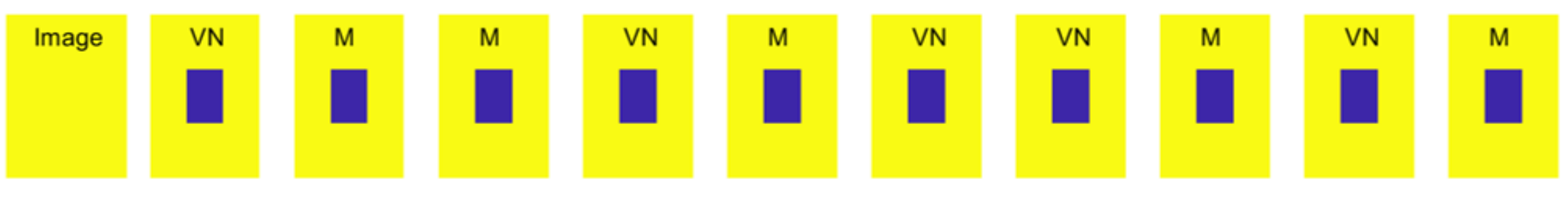}}\hspace{1em}%
  \subfloat{\includegraphics[width=0.45\textwidth]{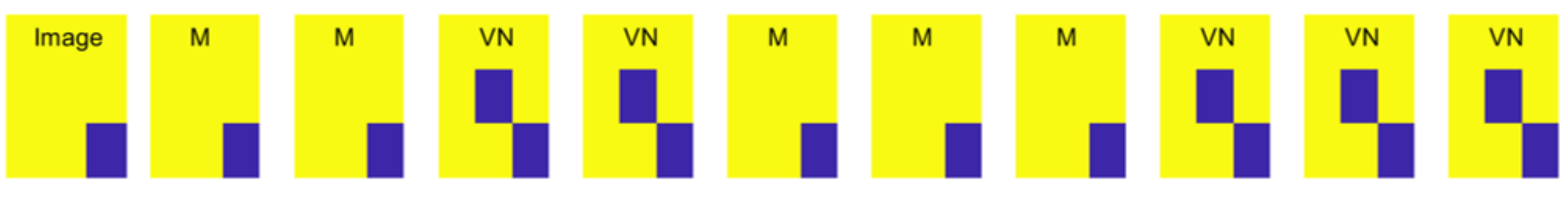}}\\
  
  \setcounter{subfigure}{0}%
  \subfloat[]{\includegraphics[width=0.45\textwidth]{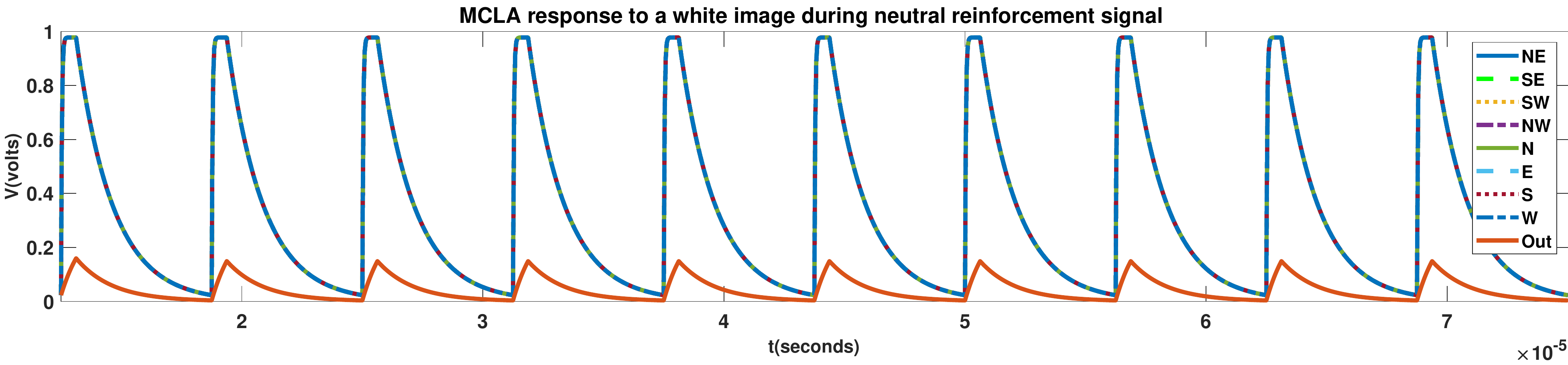}}\hspace{1em}%
  \subfloat[]{\includegraphics[width=0.45\textwidth]{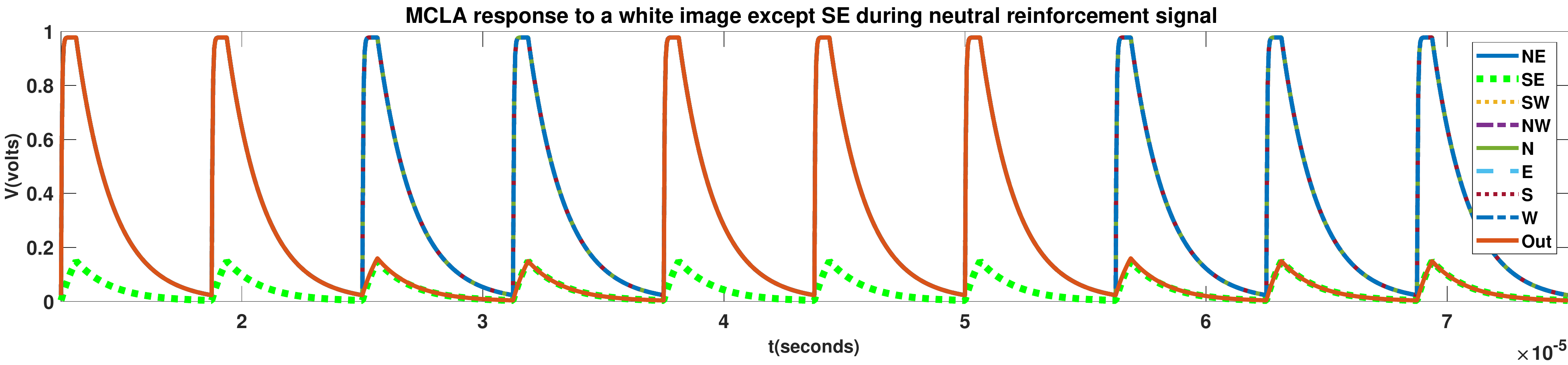}}\\
  \setcounter{subfigure}{0}%
  \subfloat{\includegraphics[width=0.45\textwidth]{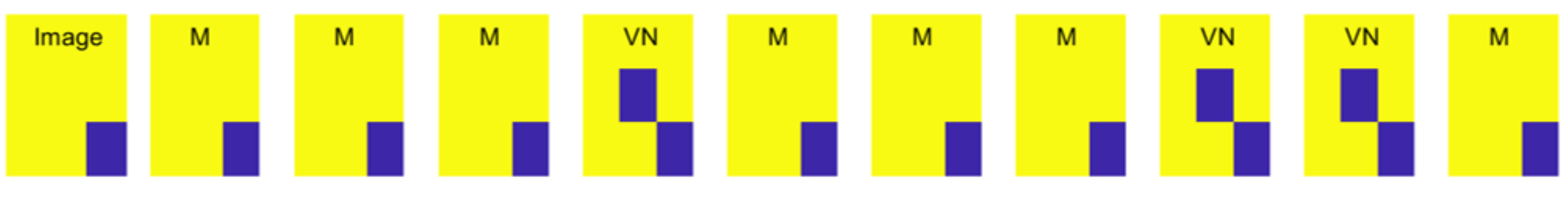}}\hspace{1em}%
  \subfloat{\includegraphics[width=0.45\textwidth]{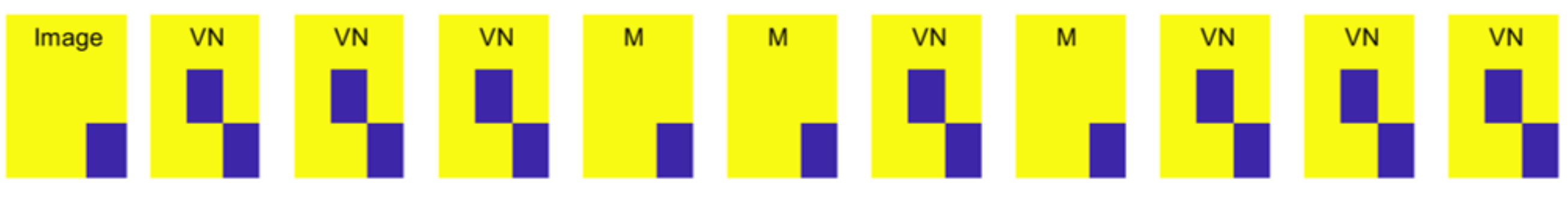}}\\
  \subfloat[]{\includegraphics[width=0.45\textwidth]{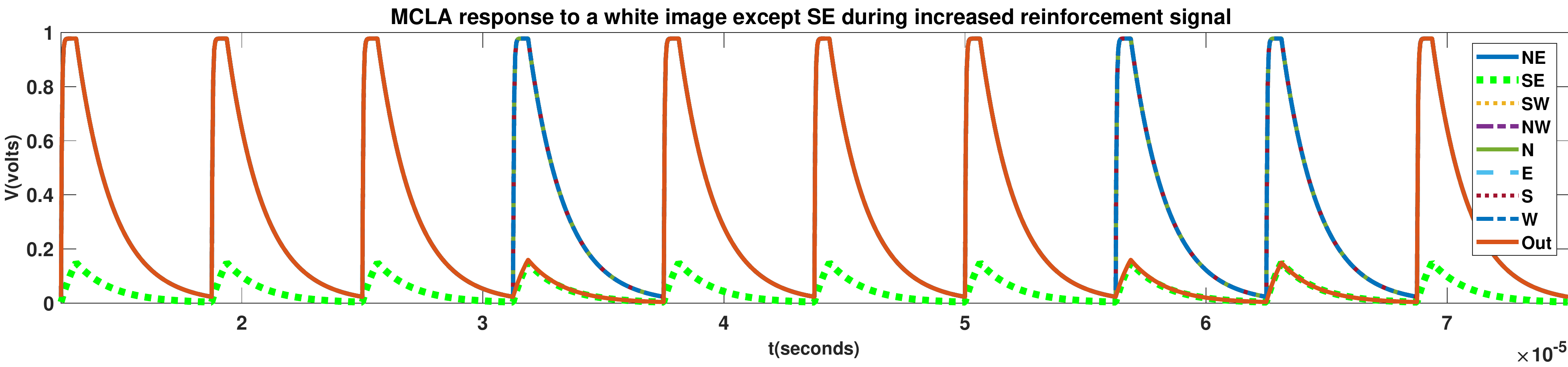}}\hspace{1em}%
  \subfloat[]{\includegraphics[width=0.45\textwidth]{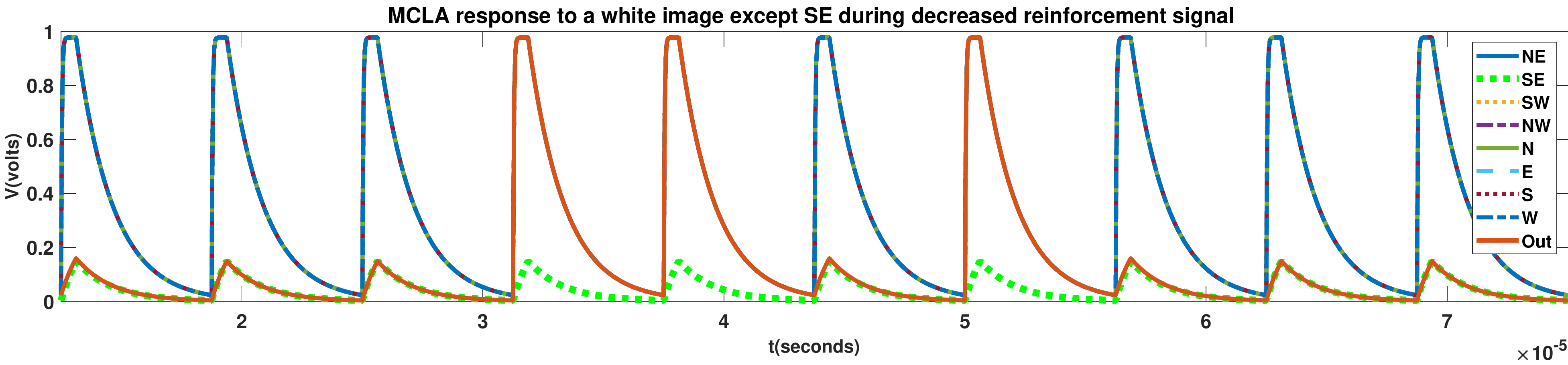}}
  
    \caption{(a) Pixel and Spike Output response to fully white image during neutral reinforcement signal. Pixel and Spike Output response to fully white image except $SE$ pixel during (b) neutral reinforcement signal, (c) increased reinforcement signal and (d) decreased reinforcement signal (color coding for every cell's state is shown in the inset).}
    \label{learning}
\end{figure*}
A circuit representation of the MLCA cell also suitable for edge detection problem is demonstrated in Fig.~\ref{Cell}.
The rule is implemented in an analog manner through Millman's theorem \cite{millman1940useful} by calculating the mean value of the input voltages at node $V_m$. When $M_1$ is in the $R_{on}$ state its resistance is equal to $R_1$ and both 4-input branches contribute equally to the calculation of the mean value at $V_m$ thus constituting a Moore's neighbourhood. On the other hand, when $M_1$ is in the $R_{off}$ state, the contribution of the diagonal neighbours to the calculation of the mean value is negligible and so we have a von Neumann neighbourhood. Switching between the two different neighbourhoods through memristor comprises the two actions of our MLCA. The stochastic selection of the neighbourhood in each time step arises from the variable switching nature of the memristor. In every time step, a $LEARNING$ voltage is applied to the memristor $M_1$, while switches $S_1$ and $S_2$ are activated. Nevertheless, this voltage is not able to switch the memristor in every write cycle, due to the memristor's variable voltage threshold owing to the device's variability. By increasing the $LEARNING$ voltage, one can increase the probability of the memristor switching and therefore manipulate the probability of the selected action. 

The environment determines whether the reinforcement $LEARNING$ voltage will be increased or decreased. The environment in our case has a specific rule, if all MLCA in the same neighbourhood choose the same action, they will be rewarded (increase in $LEARNING$ voltage), otherwise they will be penalized (decrease in $LEARNING$ voltage). To determine if all the MLCA chose the same action we examine the resistance value of the $M_1$ memristor. If all the memristors are in $R_{on}$ state, this means that all MLCAs chose the Moore's neighbourhood. Otherwise, if they are all in $R_{off}$ state, von Neumann neighbourhood was selected.

The state-memristor will be affected by the writing process only when both the rule and cell's pixel binary values are active simultaneously. So, the voltage $V_m$ in combination with the $I_{i,j}$ controls the writing process of the state memristor ($M_2$), where $I_{i,j}$ is the pixel's binary value that corresponds to the specific cell. Switch $S_3$ is by default closed and only when $V_m$ (which controls it) is above $900$mV (case where all the inputs are activated 
and therefore the state of $M_2$ should not be affected by $I_{i,j}$) opens. The writing process of the state-memristor $M_2$ is followed by the read operation. In Fig. \ref{ON_OFF_response} one can see the form of the output voltage pulses during read operation when the pixel is 
an edge (red) and when the pixel is 
not 
an edge (yellow). 

In Fig. \ref{learning} an example of edge detection in a $3\times{3}$ image, selected as small as possible for readability reasons, depicted in a $3\times{3}$ MLCA grid is demonstrated for readability reasons 
where the pixel at row $i$ and column $j$ has a yellow color if $I_{i,j}$=1, whereas it has a blue color if $I_{i,j}$=0. 
In Fig. \ref{learning}(a) a fully white (
presented here with yellow colour for readability reasons) 
image ($I_{i,j}$=1) is fed to the MLCA grid. As it can be observed in every timestep, the MLCA chooses a different neighbourhood which is denoted as VN (von Neumann) or M (Moore) accordingly due to the memristor's variability; nevertheless, output image remains the same with all cells identified as edges except the central one. The central cell remains to 0 because no matter the chosen neighbourhood action is, all the neighbouring pixels are 1, while the surrounding cells remain to 1 because no matter the chosen neighbourhood is, they are identified as edges due to the boundary conditions being 0 (grounded).  

In the following examples, all pixels of the image are set to 1 except the $SE$ pixel which is set to 0. Now, depending on the MLCA's neighbourhood selection, the central cell's output will be 1 (edge) in case of the Moore action or 0 (not an edge) in case of von Neumann action. In Fig. \ref{learning}(b) it can be observed that the probability of each action is the same due to the neutral reinforcement signal and so half of the times the central cell is identified as an edge. In this case, the von Neumann neighbourhood was chosen half the times, whereas the other half, Moore was selected. 

According to the environment rule, when all cells of the neighbourhood choose the Moore neighbourhood as an action, then an increased reinforcement signal (i.e. $LEARNING$ voltage) is applied. When an increased $LEARNING$ voltage is applied, the probability of selecting the Moore's neighbourhood is increased and consequently the cell is identified as an edge more frequently (Fig. \ref{learning}(c)). On the contrary, when all cells of the neighbourhood choose the von Neumann neighbourhood as an action, a decreased reinforcement signal (i.e. $LEARNING$ voltage) is applied. Due to the decreased $LEARNING$ voltage, the probability of selecting the von Neumann neighbourhood is increased and consequently the central cell is identified as an edge less frequently (Fig. \ref{learning}(d)). As a result, the proposed architecture of MLCA finally succeeds to deliver correct testing results for optimal detection of edges in binary images for the presented small memristive nano-crossbar grid. Further exploration of the MLCA architecture benefits when applied for bigger memristive crossbars, taking also consideration the corresponding design constraints, will be promising for the enhancement of the proposed design architecture, while application to different than image processing scientific fields will soon apply. 

\section{Conclusions}
In this paper, we exploit 
memristor device for the design and implementation of non von Neumann circuits and computing configurations, when coupled with Learning Cellular Automata principles. More specifically, we proposed a LCA architecture whose function, combined with the memristor's unique characteristics and inherent variability, is able to adapt to its environment and learn by receiving feedback from it. At first, we analyzed the basic theoretical background for the proposed MLCA and then proved its functionality with the detection of edges in image processing applications through a series of SPICE simulations for the presented circuitry.


\section*{Acknowledgement}
This work was supported in part by the European Union's H2020 research and innovation programme under the Marie Sk\l{}odowska-Curie grant agreement No 691178.

\bibliographystyle{IEEEtran}
\bibliography{biblio}

\vspace{12pt}
\end{document}